# Nonlinear Rank Scaling and Hidden Structure in NHS Expenditure Transparency Data

Animotu Mohammed,[a] Golnaz Shahtahmassebi,[b] Haroldo V. Ribeiro,[c] Jack Sutton,[a] Quentin S. Hanley[a,d]

[a]*College of Science and Engineering, University of Derby, Markeaton Street, Derby DE22 3AW, UK*
[b]*School of Science and Technology, Nottingham Trent University, Nottingham NG11 8NS, UK*
[c]*Departamento de Física, Universidade Estadual de Maringá, Maringá, PR 87020-900, Brazil*
[d]*GH and Q Services Limited, West Studios, Sheffield Road, Chesterfield S41 7LL, UK*

**\* Corresponding author:** Q. S. Hanley

**Abstract:**

A variety of transparency initiatives have been introduced by governments to reduce corruption and allow citizens to independently evaluate effectiveness and efficiency of spending. In 2010, the UK government mandated transparency for many expenditures exceeding £25,000. The resulting data is dispersed across a range of governmental organizations and presents an opportunity to understand expenditure at scale, interrogate organizational structures and develop transparency measures. Here, we focus on data from the top two layers of the National Health Service (NHS) within England, including NHS England (NHSE) and Integrated Care Boards (ICBs). As the one of the largest government run healthcare organizations in the world and potentially the sixth largest employer globally, the NHS provides a distinctive case for studying healthcare delivery, contractor dynamics, and organizational self-organization. We find that limiting transparency to larger transactions conceals a substantial share of spending from scrutiny, including most transactions. The rank-frequency distributions of suppliers, expense types, and spending categories exhibit multiple scaling regimes and these are similar to patterns observed in word frequency and urban scaling studies.

**Introduction**:

Government transparency is a broad set of concepts and policies which, among other things, reduces the asymmetry of information between governments and citizens.[1] The underlying assumption is that governments should primarily work in the interest of citizens and not in the interests of individuals within the government and openness is thought to enable this. Transparency also includes notions of accountability, quality, openness, conformance with transparency policies, financial management, human rights, whistleblower protections, and treaty verification.[2,3] A range of expert surveys and other measures are used to assess the degree of transparency achieved by a country or organization and wide range of studies have been done on the topic.[4] There are relatively few studies of transparency records. For the current work, a long time scale study of Slovenian spending is of particular interest due to evidence of self-organizing behavior, power law structures, and preferential attachment in suppliers.[5]

Financial transparency of government is of interest, particularly when resources are limited or government indebtedness is of concern. In 2010, the government of the UK initiated a requirement for financial transparency as part of a broader initiative in open government which included information about crime, the civil service, and expenditures.[6] The data released provides a window into many aspects of UK life, government, and individual and corporate safety nets. Two examples of the type of financial data provided are expenditures by the Department of Work and Pensions[7] and the Insolvency Service.[8] These reports in turn provide information about payments to companies supplying such things as IT and agency staffing needed to implement government programs. One of the key features of this data in a thresholding approach to release of expenditure: only expenditures above particular thresholds are required to be released. Relatively little has been done with the data and they present an opportunity to add to our understanding of government financial records as well as paradigms for understanding other large organizations. The data can be used to develop benchmarks that are useful for other large organizations.

One of the largest government expenditures in the UK is on health which is funded via a national insurance system and delivered by the National Health Service (NHS). The NHS is among the largest government-run healthcare systems in the world. Brazil's is the largest by number of people served while France's serves a similar sized population. As currently organized, the NHS has one of the largest overall payrolls (circa 1.4 million) of any organization worldwide. In this context, understanding how the NHS spends its money is of universal interest as a model of health care provision. Further, although there are larger private health insurance organizations by revenue, it is unlikely these providers will share similarly transparent records.

This study looks at financial transparency records of expenditure within the top two layers of the NHS: NHS England (NHSE) and the Integrated Care Boards (ICBs) used to commission health related services. The ICBs were created beginning in 2022 out of the previous clinical commissioning groups (CCGs). Below these layers are a multitude of Hospitals, Trusts, primary care organizations, hospices, and other contractors.

*Theory*

One way to present financial data is to sort by amounts transacted or by the number of transactions, leaving a rank ordered data set. An early rank ordering principle was observed by Auerbach in the context of cities where size was inversely related to rank.[9] Since then, related behavior has been observed in language,[10] income distribution[11], and gene expression.[12] Simple versions of these processes lead to power laws, which appear as linear relationships when plotted on logarithmic scales. For discrete data, the Zipf and Zipf-Mandelbrot distributions are frequently used, while continuous data is often modelled using the Pareto distribution and related distributions.[11,13] For the purposes of this discussion, we will refer to data sets that form straight or segmented lines on log-log plots as exhibiting power law ordering.

In the current context, power law ordering applies when the total amount of money transacted or the total number of transactions, $y$, received by an entity, expense type, expense area, or supplier is related to rank position, $R$, via a power law of the form[14,15]

$$y = y_0 R^\beta \qquad (1)$$

where $\beta$ will be negative for a high to low sorting. Taking the logarithm of both sides presents a linear form:

$$\log y = \log y_0 + \beta \log R \qquad (2)$$

Some of these relationships exhibit abrupt changes of slope at critical points. In rank order plots, the position of these change point(s), $r_i^*$, can be found using piecewise modelling.[16,17] Multiple change points can be treated by considering:

$$\log y = \begin{cases} \log y_0 + \beta_1 \log R & 0 < R < r_1^* \\ \log y_1 + \beta_2 \log R & r_1^* < R < r_2^* \\ \log y_2 + \beta_3 \log R & r_2^* < R < r_3^* \\ \vdots & \vdots \end{cases} \qquad (3)$$

where distinct intercepts and linear parameters capture different behaviors across intervals of rank. However, the existence of a change point or multiple change points needs to be tested using methods such as the Davies test[18] or Akaike information criterion (AIC). The change points mark transitions between different regimes and can be concave or convex and may reflect structural shifts and in the context of language a change point may represent a transition between a kernel lexicon and more extended language.[19]

Other models of rank ordered data have been described which encompass data generated by a variety of processes. One such model is a 3-parameter form based on the,[20–22]

$$y(r) = A \frac{(N-r+1)^b}{r^a} \qquad (4)$$

where $r$ is the rank, $A$ is a scale constant, $N$ is the maximum rank, $a$ determines the curvature at low values of $r$, and $b$ describes the curvature at large values of $r$. This form can be derived from a Taylor series expansion of the Yule-Simon distribution.[22] Interpretations for the parameters include mechanisms such as preferential attachment;[23] expansion and modification; and order-disorder transitions.[21] A more general form is a five parameter model encompassing the Zipf-Mandelbrot law and increased flexibility,[22,23] expressed by

$$y(r) = A \frac{(N(r+c))^b}{(N+1-r+d)^a} \quad (5)$$

where $c$ is a low rank parameter based on Mandelbrot's extension of Zipf's law and $d$ is a high rank parameter proposed by Ausloos and Cerqueti.[22] Depending on the parameterization of these equations, a range of different distributions can be invoked.

## Materials and Methods

### Data Sets.

The data sets were obtained from the web sites of NHS England and the Integrated Care Boards between 2/10/2024 and 18/11/2024. In some cases, data were missing or could not be found. The data had no standard format and were provided in a range of file types (CSV, XLSX, PDF), structures (headers, single and multiple tabs, logos, etc.), formats (particularly dates), protections, number of columns, names of columns, amount sums, and pivot tables. As these variations were not predictable, files were checked to assure preservation of date formats, header rows and column names. Where possible, the data were converted to a common format and aggregated. Data provided in PDF formats was excluded due to their varied formats and difficulty importing them reliably. This affected 4 ICBs. The data was not found for one ICB and an inquiry was sent and received no reply. This left data from 37 out of 42 ICBs and a complete set of records from NHS England. A total of 993 files were read into the data set, each representing a month of data from a reporting organization.

### Statistical Analysis.

The data were assessed and presented using R (Version 4.4.3)[24] running within R-studio (version 2024.12.1 Build 563) with packages dplyr (1.1.4),[25] readxl (1.4.5),[26] stringr (1.5.1),[27] segmented (2.1-3).[28]

**Results and Discussion**

The aggregate data set contained a total of 1,956,196 entries covering approximately 2 years between 2022 and 2024. This covers the period from when the ICBs were constituted and ended with the available data provided at the time. Although the requirement is to be transparent about transactions greater than £25,000, only 665,231 met this threshold, including 634,536 outgoing and 30,695 incoming transactions. When all records were included, there were 1,805,513 outgoing transactions worth £302,500,004,887 and 150,683 incoming transactions worth £11,060,189,124. This clearly shows that NHSE and ICB primary funding income is not fully transparent since there is an imbalance between incoming and outgoing transactions of £291.4 billion. A total of 1,290,965 under threshold transactions were provided. We note that only some (34 out of the 37) ICBs reported below threshold transactions and that the number below threshold transactions varied from a few (NHS Beds, Luton & MK ICB) to 105,182 (NHS West Yorkshire ICB).

**Variable Transparency and Transaction Amount Histograms.**

To investigate the characteristics of the transparency data set, a histogram of the amounts of all outgoing transactions was constructed (Figure 1a). This revealed a possible bimodal structure with a valley around £25k. Upon investigation this was found to be an artifact of different approaches to meeting the transparency requirement. Some of the commissioning bodies cut the data provided exactly at £25k (Figure 1b), while others were more expansive and appear to have released all outgoing transactions (Figures 1c and 1d). This provided an opportunity to compute two transparency indices: i) an index of excess transaction number transparency (number of total records divided by the number of transactions ≥ £25k for each commissioning entity) and ii) an index of excess financial amount transparency (total amount divided by total amount in transactions ≥ £25k). In turn these can be viewed by income, expenditure, and all records.

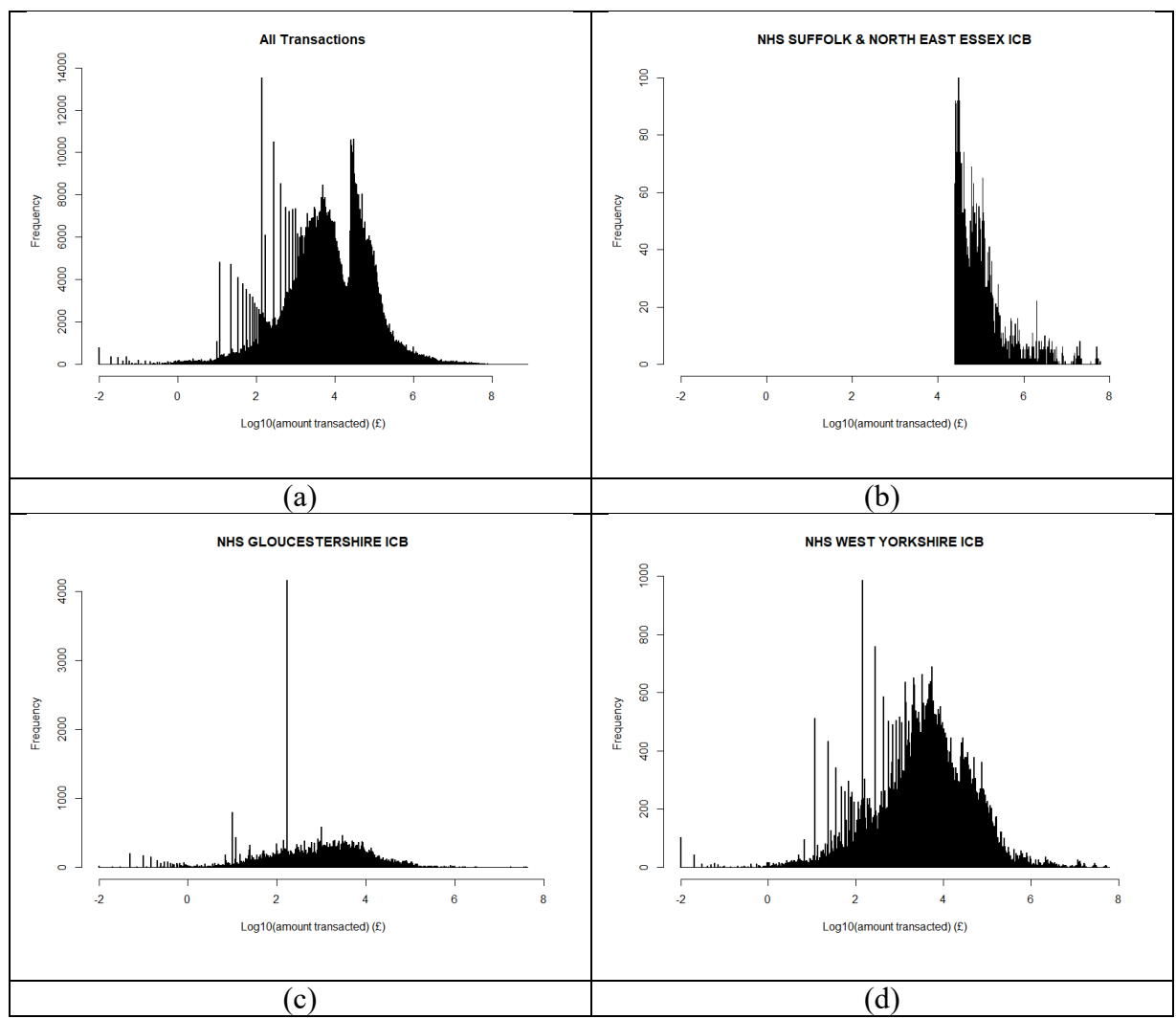

Figure 1: Histograms of expenditure showing variability in interpretation of the >£25k reporting requirement. Histograms include: (a) all transaction amounts, (b) amounts from an ICB adhering strictly to the £25k mandate, (c) amounts from a high excess transparency ICB, and (d) amounts from an ICB with lower excess transparency.

Investigation of the 6 resulting transparency indices found considerable variation across the entities presented in the data set (Figure 2). For *incoming* transactions, 48 entities were represented with excess transaction number transparency ranging from 1.00 (9 entities) to 29.86 (NHS Gloucestershire ICB) and amount transparency from 1.00 (4 entities) to 1.60 (NHS Devon ICB). For *expenditure*, 50 entities had excess transaction number transparency from 1.00 (11 entities) to 13.06 (NHS Gloucestershire ICB) and amount transparency from 1.00 (11 entities) to 1.09 (NHS Gloucestershire). For *all* transactions, the excess transaction index ranged from 1.00 (2 entities) to 15.02 (NHS Gloucestershire) while the excess amount index ranged from 0.85 to 1.03. In the last case, excess amount transparency index values below 1 are particularly interesting because they only occur when there is significant income.

Overall, when evaluating large similar organizations working to identical threshold guidance, these transparency indices indicate that it is impossible to estimate how much information will be lost when a reporting threshold is set. In this data set, the public's view of amounts would have been off by as much as 37.5% (income NHS Devon ICB) and by a factor of almost 30 for transactions (income NHS Gloucestershire ICB). From the point of view of transparency, there is value in providing all transactions and modern data science makes working at this scale relatively straightforward. Further, the below threshold transactions are important. For example, the highest point in three of the histograms (Figure 1, panels a, c, and d) corresponds to a transaction value of £140. This unit of expenditure is generated when a learning disability health check is done by a general practitioner (GP). These transactions and the associated amounts are invisible in the remaining histogram (Figure 1, panel b). This example is just a single case of small payments made to GPs. Payments for Delegated GP services represent 996,249 (77.2%) of the 1,290,965 under-threshold transactions, all of which would be invisible in a strict > £25k view.

Outside of the transparency indices, the variable number of entities appearing in the panels of figure 2 highlights the variability of individual ICBs managing reorganization and how they account for commissioning Support Units (CSUs). Four ICBs show CSUs as suppliers, while others consider them as internal entities that appear without any visible incoming transactions. In the former case, expenditures on the CSUs are visible as a transaction to a supplier (the CSU), while, in the other case, the existence of the CSU is only known because it is spending money.

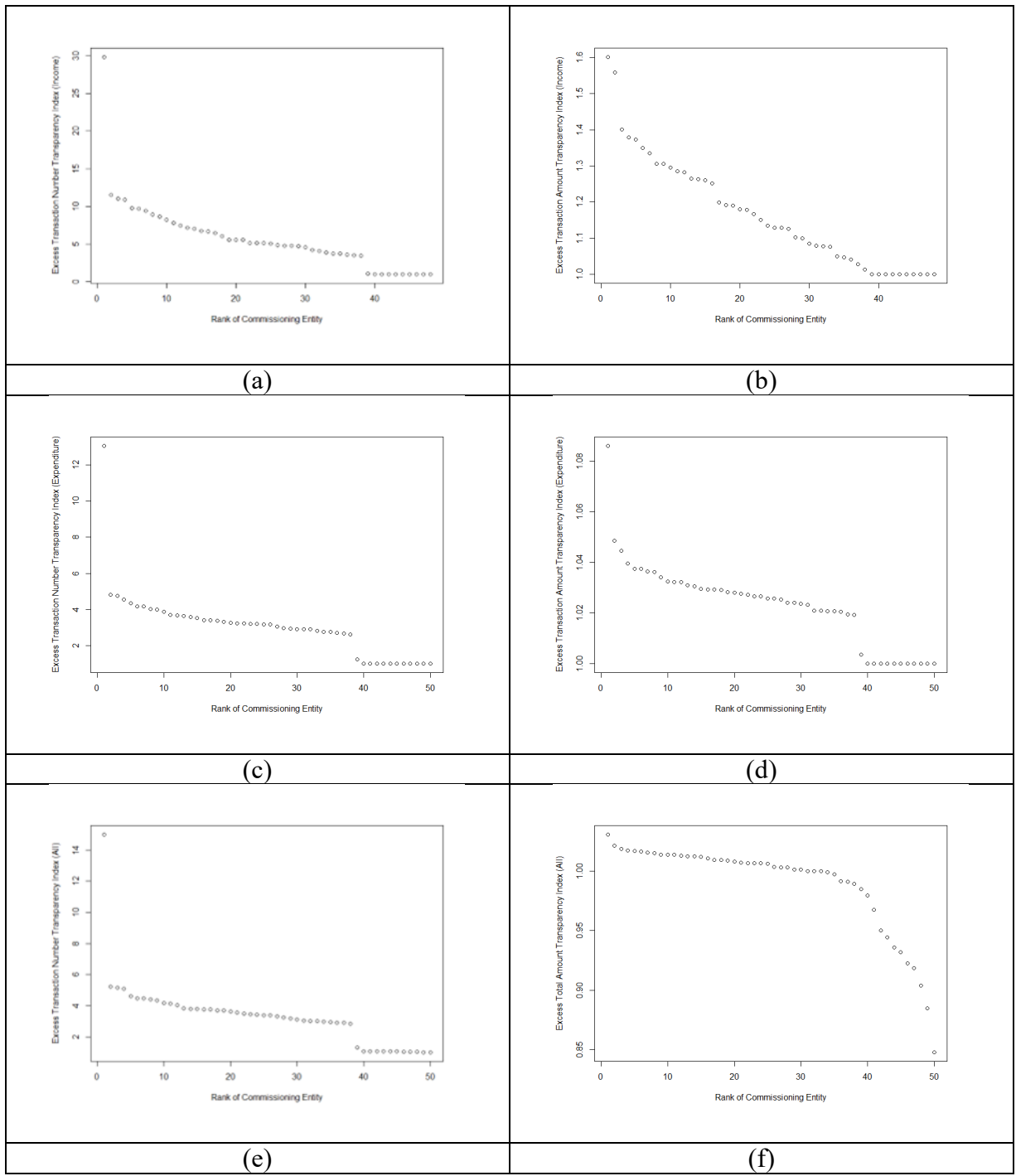

Figure 2: Excess Transparency Index Values. Rank order presentation of excess transparency index for income (a and b), expenditure (c, and d) and for all transactions (e and f). Indices are shown for transactions (a, c, and e) and amounts (b, d, and f).

**Supplier scaling**.

A total of 17,062 suppliers were represented in the data. A rank order plot of the number of transactions generated by each supplier conformed to a segmented Zipf's law type relationship with a changepoint near the 5900$^{th}$ ranked supplier (Figure 3a). These large transaction volume suppliers represented 34.6% of suppliers and generated 1,754,524 (89.7%) transactions of which circa 1.1 million were for GP services. The remaining 11,152 suppliers generated only 201,606 transactions. To our knowledge, this type of behavior has not been reported previously in the context of large-scale organizations although piecewise analysis of rank order data has been reported in studies of languages[10,19] and corruption index ranking of countries.[29]

A ranked plot of aggregate value of the transactions of each supplier also revealed a segmented relationship with log(rank) (Figure 3b). The change point was observed circa ranked supplier number 221 and represents the approximate location of the transition between the recipients of nearly all funds and the remaining suppliers. The first 221 (1.3%) suppliers received £237,795,574,537 (81.6%) while the remaining 16,841 (98.7%) received £53,642,034,354 (18.4%). The 221 suppliers receiving the greatest amounts were mostly other NHS and government organizations (e.g. hospitals, trusts, councils, and HM Revenue and Customs) with 10 limited companies and one community interest corporation (CIC) also present. The extreme concentration of transaction aggregate amounts in a few suppliers, most of which are sub-units of the same organization is noteworthy. There is some overlap between the top suppliers in terms of amounts and the top generators of transactions. For example, HMRC appears at number 4 in terms of transaction numbers and 112 in terms of total amounts. Alternative presentations indicate that this presentation masks additional information for amounts (figure 3c). Of interest is the sharp downward break £25k near supplier 16,284 indicative of a truncation of information about small suppliers. This strongly suggests the true number of suppliers is much larger than found in the transparency data. This reinforces the value of providing all transactions.

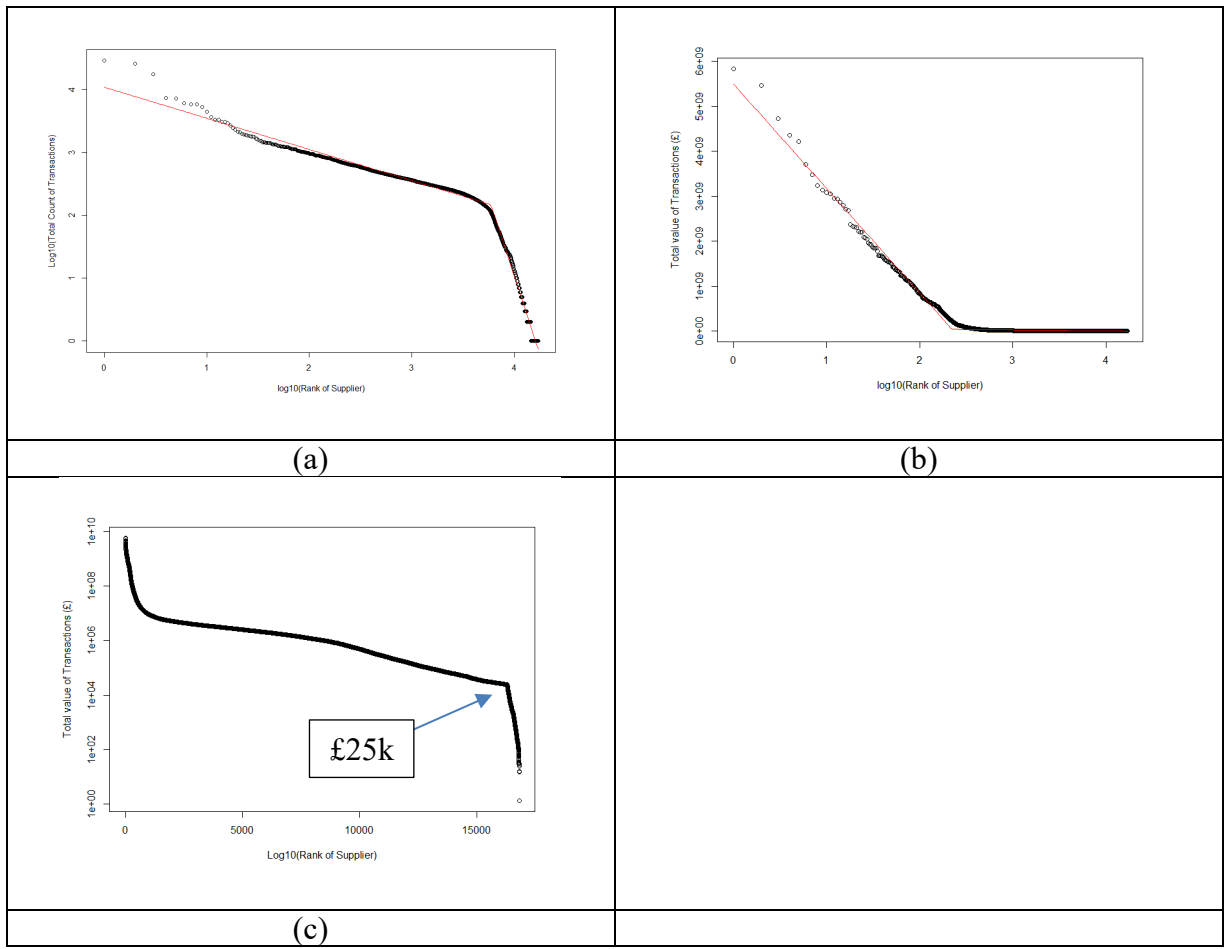

Figure 3: Rank order of suppliers: (a) log-log presentation of the transactions generated with piecewise linear fit, (b) log rank presentation of amounts transacted by suppliers with piecewise linear fit, (c) log amount presentation of suppliers highlighting the partial truncation of suppliers at £25k.

**Major Costs and Scaling of Expense Types and Expense Areas**

The previous views of the NHS data set were not specific to healthcare beyond noting that threshold limits would mask most spending on GP services. To better understand healthcare spending, two additional columns in the data set were examined: *expense type* and *expense area*. There were 1117 different *expense types* and 316 *expense areas* in the data. Due to the number of *expense types* and *expense areas* a complete listing is not presented; however, by amount transacted, the top 5 *expense types* and *expense areas* accounted for 70.77% and 64.74% of funds, respectively (Table 1). By number of transactions, the top 5 *expense types* accounted for only 18.84% of the number of transactions but, by *expense area*, delegated GP services made up 64.26% alone (Table 2).

Table 1: Amount transacted by expense type and area.

| Rank | Expense Type | Amount (£) | Percent of total |
|---|---|---|---|
| 1 | HCARE SRV REC FDTN TRUST-CONTRACT BASELINE | 67,786,625,246 | 23.88 |
| 2 | SERVICES FROM OTHER NHS BODIES | 53,484,414,605 | 18.84 |
| 3 | HCARE SRV REC NHS TRUST-CONTRACT BASELINE | 39,337,891,061 | 13.86 |
| 4 | HLTHCRE-FOUNDATION TRSTS | 28,734,651,447 | 10.12 |
| 5 | HCARE SRV REC FDTN TRUST-NON CONTRACT | 11,553,867,473 | 4.07 |

| Rank | Expense Area | Amount (£) | Percent of total |
|---|---|---|---|
| 1 | ACUTE SERVICES A | 110,727,116,647 | 39.76 |
| 2 | SPECIALISED COMMISSIONING | 24,658,847,730 | 8.85 |
| 3 | DELEGATED GP | 18,759,105,838 | 6.74 |
| 4 | MENTAL HEALTH SERVICES A | 14,211,542,814 | 5.10 |
| 5 | NHS ENGLAND RUNNING COSTS | 1,194,518,8422 | 4.29 |

Table 2: Number of transactions by expense type and area. Note: GMS refers to the General Medical Services contract which is the framework for delivery of most GP services.

| Rank | Expense Type | Number of Transactions | Percent of total |
|---|---|---|---|
| 1 | C&M-GMS GLOBAL SUM | 95530 | 5.07 |
| 2 | C&M-GMS QOF ASPIRATION | 79070 | 4.20 |
| 3 | C&M-GMS PCN DES PARTICIPATION | 74377 | 3.95 |
| 4 | SERVICES FROM OTHER NHS BODIES | 56445 | 2.99 |
| 5 | CLINICAL&MEDICAL-INDEPENDENT SECTOR | 49664 | 2.64 |

| Rank | Expense Area | Number of Transactions | Percent of total |
|---|---|---|---|
| 1 | DELEGATED GP | 1199686 | 64.26 |
| 2 | LOCAL INCENTIVE SCHEMES | 101974 | 5.46 |

| 3 | CHC ADULT FULLY FUNDED | 52702 | 2.82 |
| 4 | DELEGATED OPHTHALMIC | 41089 | 2.20 |
| 5 | ACUTE SERVICES A | 37846 | 2.03 |

When viewed as a whole, the ranked expense types and areas conformed well to existing 3- and 5-parameter models (Figure 4). When considering amounts, expense type (Figure 4a) fit well to a 3-parameter model (equation 3) while expense area (Figure 4b) conformed better to a 4-parameter model (equation 4 with the $c$ parameter = 0). Transaction numbers required all five parameters for expense type (Figure 4c) and 4 parameters (equation 4 $c$ parameter = 0) for expense areas (Figure 4d). A key area where this type of model has found utility is in analysis of word frequencies in a range of contexts. The emergence of similar behavior in NHS financial transparency data suggests that the terms used to designate expenditure types function as descriptive vocabularies. The amounts transacted and the number of transactions serve as vocabulary usage metrics. This view of large-scale financial accounting data connects the transparency data to mathematical frameworks developed in studies of language, urban systems, and economic inequality.

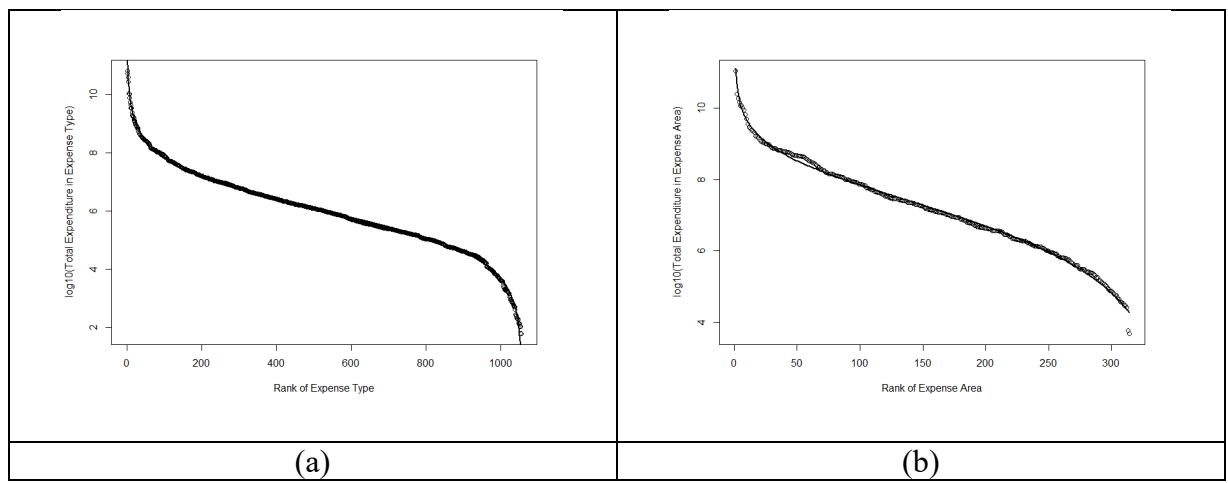

| (a) | (b) |

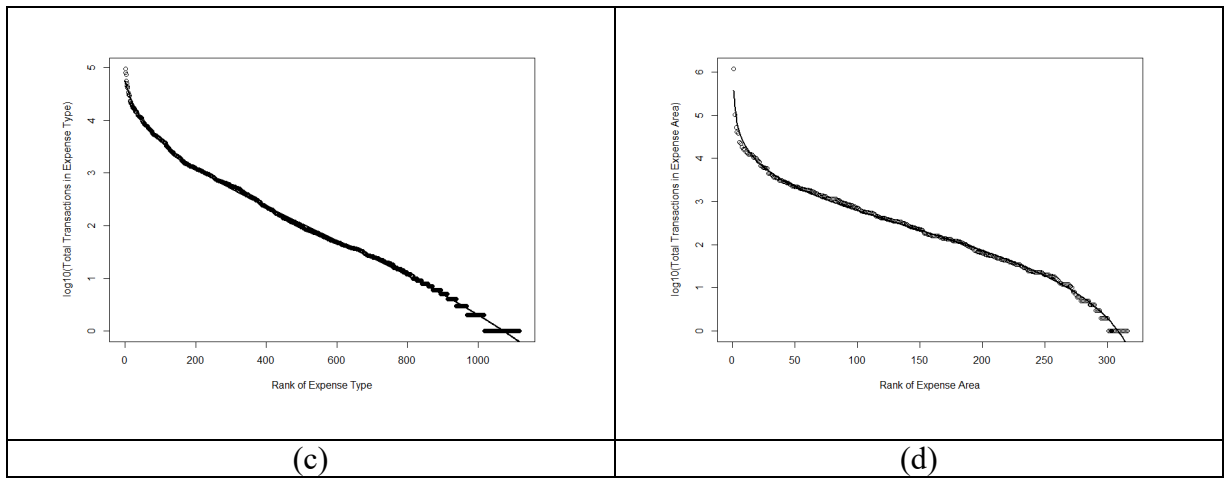

Figure 4: Rank order behavior of expense type (a and c) and expense area (b and d) by total amount in GBP (a and b) and number of transactions (c and d).

**Conclusion**

Although financial transparency data has been mandated for many years in the UK, efforts to systematically collate and analyze this information remain limited. The data set examined here was subject to a requirement to release expenditures over £25k; however, the implementation of this mandate is inconsistent across institutions. There clearly is no restriction on releasing more data than required and the NHS organizations providing information below the £25k requirement make clear that imposing a threshold limits transparency. In the absence of all the data, neither the number nor the value of transactions can be readily estimated. This is seen in a histogram of all transactions (figure 1), reinforced in the range of excess transparency indices (Figure 2), and suggested by partial truncation of suppliers (Figure 3). There is incomplete transparency on the income side to confirm that outgoing payments from NHSE balance incoming payments to the ICBs. From the point of view of transparency, the data provided make it impossible to get a full picture. While there is no direct evidence, the possibility that large-scale inefficiencies or corruption could be obscured beneath the £25,000 reporting threshold cannot be dismissed. As such, full transparency really needs a complete listing of income and expenditure. The mandate for transparency shaping the data analyzed here is approximately 15 years old. In that time, data science has advanced significantly, enabling individual citizens with modest computational resources to process the financial transactions of major government programs, such as the NHS, conduct comparative analyses, and draw informed conclusions.

The behavior of health-related spending follows mathematical patterns observed in a range of other contexts. This suggests that healthcare descriptors (e.g. expense types and expense areas) can be treated as a lexicon producing usage metrics (transaction amounts and numbers) and there is good evidence for a kernel lexicon and a more extended language in NHS suppliers. Viewed this way, these spending patterns can be analyzed analogously to textual data.

**Acknowledgements**

Pilot work on this project was done as part of MSc Data Science projects at the University of Derby. This work was supported by internal University of Derby funds.